\documentclass[aps,eqsecnum,nofootinbib,showpacs,twocolumn]{revtex4}
\usepackage[dvips]{graphicx}  
\usepackage{amsmath,amsfonts}
\usepackage{bm}
\usepackage{color}  
\begin{document}
\title{A Confinement Potential for Leptons \\
and \\
Their Tunneling Effects into Extra Dimensions}
\author{Akira Kokado}
\email{kokado@kobe-kiu.ac.jp}
\affiliation{Kobe International University, Kobe 658-0032, Japan}
\author{Takesi Saito}
\email{tsaito@k7.dion.ne.jp}
\affiliation{Department of Physics, Kwansei Gakuin University,
Sanda 669-1337, Japan}
\date{\today}
\begin{abstract}
Considering the five-dimensional warped spacetime $AdS_5$ with  the $D3$-brane, we derive a potential 
in the fifth dimension, according to which ordinary particles are initially confined on the  $D3$-brane. 
It is estimated, however, that the lightest 
neutrino with mass $m_1$ is tunneling away into the extra dimension. Hence there is a possibility that no neutrinos 
with mass $m_1$ exist in cosmic background neutrinos, but surviving neutrinos are those with heavier masses $m_2$ and $m_3$. 
The other possibilities are also discussed.
\end{abstract}
\pacs{14.60.Lm, 14.60.-z, 04.50.-h, 11.25.Uv}
\maketitle
\section{Introduction}\label{sec:intro}
%
In the present paper we would like to discuss some behaviors of particles in extra dimensions. We take the so-called braneworld picture. 
In this scenario, ordinary matter is trapped in a three-dimensional space, called the  $D3$-brane, embedded in a higher dimensional space. 
This idea must be contrasted with the traditional view of extra dimensions, the Kaluza-Klein picture, where matter fields live everywhere 
in compact extra dimensions. Any such higher dimensional field can be described as an infinite collection of four-dimensional fields, 
the so-called KK modes, with masses depending on the size of the extra dimensions. Non-observation of KK modes in the present collider 
experiments suggests the size to be very small. Hence we do not take such a picture here. \\
\indent We focus our attention on the problem about what kinds of particles can move to the extra dimension. We confine ourselves to 
consider only leptons, because this possibility may be large for particles with small masses. We consider the five-dimensional warped spacetime $AdS_5$  
with the $D3$-brane \cite{ref:Randall}-\cite{ref:Langlois}. We derive a potential in the fifth dimension, according to which ordinary leptons are 
initially confined on the the $D3$-brane. Then we would like to discuss whether these particles on the brane can move to the fifth dimension 
by something like a tunneling effect \cite{ref:Photon_ex}.\\
\indent Main results are the following: Initial numbers of neutrinos will be found to decrease because of escapes into the fifth dimension by tunneling. 
From the data of solar neutrinos \cite{ref:Mikheyev} we estimate the half-life times of neutrinos with masses  $m_1, m_2,$ and $m_3$, respectively. The lightest neutrino 
with $m_1$ is only tunneling away into the extra dimension. Hence there is a possibility that no neutrinos with mass $m_1$ exist in cosmic background neutrinos, 
but surviving neutrinos are those with heavier masses $m_2$ and $m_3$. The other possibilities are also discussed. \\
\indent In Sec.\ref{sec:2} we derive such a potential in the extra dimension. In Sec.\ref{sec:3} the tunneling effect is discussed. 
In Sec.\ref{sec:4} we derive the dispersion relation in the bulk. The final section is devoted to concluding remarks.
  
\section{A potential in the extra dimension}\label{sec:2}
Let us consider a five-dimensional spacetime with three-dimensional isotropy and homogeneity metric 
\begin{align}
 &  ds^2 = e^{-2\eta |y|}\big(dt^2 - d\vec{r}^{~2} \big) - dy^2~,
  \label{eq:metric}
\end{align}
where the brane-universe is located at $y=0$ and is spatially flat \cite{ref:Randall}-\cite{ref:Langlois}. \\
\indent This metric has been obtained from the five-dimensional Einstein equations
\begin{align}
 &  G_{AB} + \Lambda g_{AB} = \kappa ^2 T_{AB}.
  \label{eq:Einstein_Eq}
\end{align}
where $\Lambda $ is a cosmological constant in the bulk. Here the energy-momentum tensor $T_{AB}$ is decomposed 
into a bulk and a brane. The former is assumed to vanish, while the latter has a form of $T_{AB}=S_{AB}\delta (y)$, 
which is taken into account of only the brane tension. The absolute value $|y|$ in the exponential factor comes 
from this delta function $\delta (y)$. The coefficient $\eta $ is related to the cosmological constant \cite{ref:Randall}-\cite{ref:Langlois}. \\
\indent  Now we would like to discuss whether the ordinary leptons on the brane can move along the geodesic line of Eq. (\ref{eq:metric}) 
by something like a tunneling effect \cite{ref:Photon_ex}. In order to see this we consider the action
\begin{align}
 &  I = \int d\tau \mathcal{L}~,
  \label{eq:Lagrangian} \\
 & \mathcal{L} = -m\sqrt{e^{-2\eta |y|}\Big\{\big(\frac{dt}{d\tau }\big)^2 - \big(\frac{d\vec{r}}{d\tau }\big)^2 \Big\} - \big(\frac{dy}{d\tau }\big)^2 }~.
  \nonumber
\end{align}
which reduces to, by choosing $\tau = t$, 
\begin{align}
 & I = \int dt L~,
  \label{eq:action2} \\
 &  L= -m\int dt  \sqrt{e^{-2\eta |y|}\big\{1 - \dot{\vec{r}}^{\ 2} \big\} - \dot{y}^2 }~.
  \nonumber \\
 & \dot{\vec{r}}= \frac{d\vec{r}}{dt}~, \quad \dot{y}=\frac{dy}{dt}~,
 \nonumber 
\end{align}
where $m$ is a parameter with the mass-dimension.
We fix its value to be the four-dimensional particle mass, when $y=\dot{y}=0$. \\
\indent Conjugate momenta are given as
\begin{align}
 & \vec{p} = \frac{\partial L}{d\dot{\vec{r}}} = \frac{m\dot{\vec{r}} e^{-2\eta |y|}}{\sqrt{e^{-2\eta |y|}(1-{\dot{\vec{r}}}^2) - {\dot{y}}^2}}~,
 \label{eq:def_p} \\
 & p_y = \frac{\partial L}{d\dot{y}} = \frac{m\dot{y}}{\sqrt{e^{-2\eta |y|}(1-{\dot{\vec{r}}}^2) - {\dot{y}}^2}}~.
 \label{eq:def_py} 
\end{align}
The Hamiltonian is
\begin{align}
 & H= \vec{p}\cdot \dot{\vec{r}}+p_y\dot{y}-L
 \label{eq:def_Hamiltonian} \\
 & = \frac{m e^{-2\eta |y|}}{\sqrt{e^{-2\eta |y|}(1-{\dot{\vec{r}}}^2) - {\dot{y}}^2}}~.
 \nonumber
\end{align}
Eliminating $\dot{r}$ and $\dot{r}$ from Eqs.(\ref{eq:def_p})-(\ref{eq:def_Hamiltonian}), we have
\begin{align}
 & H = \sqrt {\vec{p}^{\ 2} + e^{-2\eta |y|}({p_y}^2 + m^2)}~.
 \label{eq:Hamiltonian2}
\end{align}
Hamilton's equations of motion are
\begin{align}
 & \dot{\vec{r}} = \frac{\partial H}{\partial \vec{p}}= \frac{\vec{p}}{H}~,
 \nonumber \\
 & \dot{\vec{p}} = - \frac{\partial H}{\partial \vec{r}} = 0~,
 \nonumber \\
 & \dot{y} = \frac{\partial H}{\partial p_y} = \frac{p_y}{H}e^{-2\eta |y|}~,
 \label{eq:EOM0} \\
 & \dot{p_y} = - \frac{\partial H}{\partial y} = \pm \frac{k e^{-2\eta |y|}}{H}({p_y}^2 + m^2)~,
 \nonumber \\  
 & \quad (+ \mbox{ for } y>0, \ - \mbox{ for } y<0)
 \nonumber
\end{align}
From Eqs.(\ref{eq:Hamiltonian2})-(\ref{eq:EOM0}) we get
\begin{align}
 & \vec{p} = \vec{c_1} = \mbox{const.}~,
 \label{eq:sol_p} \\
 & H = \sqrt{ \vec{c_1}^2 + e^{-2\eta |y|}(p_y^2 + m^2)} = c_0 = const.~,
  \label{eq:sol_H} \\
 & \dot{\vec{r}} = \mbox{const.}~.
  \label{eq:sol_r}
\end{align}
An equation for the extra dimension is obtained from Eq.(\ref{eq:sol_H}) 
as
\begin{align}
 & \frac{p_y^{\ 2}}{2m} + U(y) = E_y =0~,
 \label{eq:Energy_y1}
\end{align}
where
\begin{align}
 & U(y) = \frac{m}{2}\big[1 - q e^{2\eta |y|}\big]~,
 \label{eq:Potental_U} \\
 &  q\equiv \big(c_0^2 - \vec{c}_1^{\ 2}\big)/m^2~.
 \nonumber
\end{align}
Here $p_y^{\ 2}/2m$ and $U(y)$ can be regarded as a kinetic energy and a potential of the particle in the extra dimension, respectively, with the total energy $E_y=0$.\\
\indent Let us suppose that the brane at $y = 0$ is initially an extended object with a thin width, $-\epsilon < y < \epsilon $.  In order that the particle with zero-energy is confined in this region, the potential should be positive at  $y=\pm \epsilon$. So, we require the condition                                             
\begin{align}
 & U(\pm\epsilon ) \cong   \frac{m}{2}(1-q) \equiv Q > 0~,
 \label{eq:def_U}
\end{align}
where $1-q=-p_y^{\ 2}(\pm\epsilon )/m^2 >0$ is fixed to be a positive constant. The other case, $i.e.$, 
$p_y^{\ 2}(\pm\epsilon )\geq 0$, the particle with zero-energy can not be confined in this region, 
$-\epsilon < y < +\epsilon $.  So, we do not consider such a case. In the region, $-\epsilon < y < \epsilon $,  the potential $U(y)$ is assumed to be zero. \\
%
%
\begin{figure}
\label{fig1}
 \begin{center}
  \includegraphics[width=6.5truecm,clip]{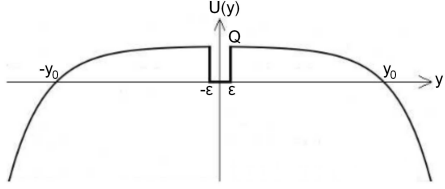}
  \caption{ 
  A plot of the potential $U(y)$ as a function of  $y$. The $D3$-brane at $y = 0$ is initially an extended object with a thin width, $-\epsilon < y <\epsilon $. 
  } 
 \end{center}
\end{figure}
%
%
\indent The potential is symmetrical as depicted in Fig.1. The curve crosses the $y$-axis at $y=\pm y_0$, satisfying $\exp{(-2\eta |y_0|)}=q$. Particles are initially 
confined in the $D$-brane. Hence each of them is supposed to be in a virtually bound state with some energy $E_0$. Some kinds of particles could escape 
from the inside to the outside of the potential by the tunneling effect. In the next section we calculate this tunneling probability$P$. The reflection probability $R$ is, 
of course, given by $R=1-P$.
\section{Tunneling effects} \label{sec:3}
Initially let a particle be confined inside the D-brane with the binding energy $E_0$. This means that the total energy should be replaced by
\begin{align}
 & \frac{p_y^{\ 2}}{2m} +U(y) = E_0~,
 \label{eq3:EOF_py}
\end{align}
\indent The tunneling probability is then given by

\begin{align}
 & P \cong  \exp{\Big[ -\frac{2}{\hbar } \int _{\epsilon }^{y_1} dy \sqrt{2m\big(U(y)-E_0\big)}\Big]}~.
 \label{eq:TunnelingA}
\end{align}
in the WKB approximation. Here $y_1$ is given by $U(y_1)=E_0$. The integral is carried out exactly as follows: 
\begin{align}
 & I = \frac{2}{\hbar }\int _{0}^{y_0} dy\sqrt{2m\big(U(y)-E_0\big)}
 \nonumber \\
 & =\frac{2}{\hbar }\int _{0}^{y_0} dy\sqrt{2m'U'(y)}~,
 \label{eq:TunnelingA2}
\end{align}
where
\begin{align}
 & U'(y) = \frac{m'}{2}\big(1-q'\exp{(2\eta y)}\big)~,
  \label{eq3:def_U'} \\
 & m' = m\sqrt{1-\frac{2E_0}{m}}~,
  \label{eq3:def_m'} \\
 & q' = \frac{q}{1-\frac{2E_0}{m}}~,
  \label{eq3:def_q'}
\end{align}
and $\epsilon \to 0$. Then it follows that
\begin{align}
 & I = \frac{2m'}{\hbar \eta }f(q')
 \label{eq3:cal_I}
\end{align}
with
\begin{align}
 & f(q') = -\sqrt{1-q'} + \frac{1}{2}\ln\frac{{2\sqrt{1-q'}+2-q'}}{q'}~. 
 \label{eq3:def_f} 
\end{align}
Here $q'$ does not depend on particle masses \cite{ref:mass_no_dependence}. This property is nothing but the equivalence principle of General Relativity. Substituting the result (\ref{eq3:cal_I}) into Eq.(\ref{eq:TunnelingA}) we get
\begin{align}
 & P \cong \exp{\Big[-\frac{2m'}{\hbar \eta }f(q') \Big]}~.
 \label{eq:TunnelingA3}
\end{align}
\indent From Eq.(\ref{eq:TunnelingA3}) we see that the tunneling probability depends sharply on the mass parameter $m'$. For  $\eta ^{-1}=10^{-2}$cm \cite{ref:Langlois}, 
we have tunneling probabilities for leptons other than neutrinos listed in Table~\ref{Table1}. Here we have assumed to be $2E_0/m << 1$, hence $m' \cong m$ 
and $q' \cong q$. The tunneling probabilities seem to be very small, actually regarded so as to be zeros because of large masses of leptons, 
if $f(q)$ is not so small. In fact we will see later a fact that $f(q)$ should take values larger than 10.8. \\ 
%
%
\begin{table}[h]
\begin{center}
\begin{tabular}{cc}
\\
\hline
\hline
$m_{e}=0.5$ MeV$/c^2$~~  &  $P_{e}=\exp{(-0.5\times 10^9 f(q))}$~~  \\
$m_{\mu }=106$ MeV$/c^2$~~  &  $P_{\mu }=\exp{(-1.1\times 10^{11} f(q))}$~~  \\
$m_{\tau }=1.7$ GeV$/c^2$~~  &  $P_{\tau }=\exp{(-1.7\times 10^{12} f(q))}$~~  \\
\hline
\hline
\end{tabular}
\caption{ Tunneling Probabilities for leptons.}
\label{Table1}
\end{center}
\end{table}
%
%
\indent As for masses of three mass-eigenstates of neutrinos $\nu _{1}, \nu _{2}$, and $\nu _{3}$, we take $m_1 = 0.001$eV/$c^2$, $m_2 = 0.01$eV/$c^2$, and $m_3 = 0.05$eV/$c^2$ \cite{ref:Mikheyev}. 
Three kinds of flavor neutrinos $\nu _{e}, \nu _{\mu },$ and $\nu _{\tau }$  are composed of $\nu _{1}, \nu _{2},$ and $\nu _{3}$. The tunneling probabilities for $\nu _{1}, \nu _{2},$ and $\nu _{3}$ are also given by Eq.(\ref{eq:TunnelingA3}). Under the same assumption as $2E_0/m << 1$, hence $m'\cong m$, and $q'\cong q$ we have 
\begin{align}
 & P_1 \cong \exp{[-f(q)]} 
 \nonumber  \\
 & = \exp{(-10.8)} \cong  2.0\times 10^{-5}~,
  \label{eq3:Prob_P1}
\end{align}
for $m_1$ = 0.001 eV/$c^2$, $\eta $= 100$ cm^{-1}$, and $f(2.25\times 10^{-10})=10.8$ \cite{ref:Mikheyev}. 
The zero point of the potential becomes $y_0=0.11$ cm, which is given by the equation $q=\exp{(-2\eta y_0)}$. 
The non-zero value of the tunneling probability is a consequence of the small mass of the neutrino compared with other leptons. It may be convenient to use the half-life time, 
because the number of neutrinos in the $D$-brane decreases by the tunneling effect. The half-life time $T_1$ for $m_1$ is defined by
\begin{align}
 & T_1 = \frac{\ln{2}}{P_1} = 9.6~h~,
 \label{eq3:Half_time_P1} 
\end{align}
In the same condition as $f(q)=10.8$, we have $T_2=10^{39}$ yr and $T_3=10^{226}$ yr for $m_2$ and $m_3$, respectively.  Note that those half-life times $T_2$ and $T_3$  are longer than the age of our universe $10^{10} yr$. In Table~\ref{Table2} we have listed the half-life times for various values of $f(q)$. \\
%
%
\begin{table}[h]
\begin{center}
\begin{tabular}{ccccc}
\\
\hline
\hline
$f(q)$~~ & $q$~~ & $T_1$~~ & $T_2$~~ & $T_3$~~ \\
\hline
$10.8$~~  &  $10^{-10}$~~ & 9.6 h~~ & $10^{39}$ yr~~ & $10^{226}$ yr~~ \\
$29.8$~~  &  $10^{-27}$~~ & $10^{5}$ yr~~ & $10^{121}$ yr~~ & $10^{640}$ yr~~ \\
$41.0$~~  &  $10^{-36}$~~ & $10^{10}$ yr~~ & $10^{170}$ yr~~ & $10^{882}$ yr~~ \\
\hline
\hline
\end{tabular}
\caption{ Half-life times for three kinds of neutrinos against $f(q)$}
\label{Table2}
\end{center}
\end{table}
%
%
\indent The solar neutrinos reach the earth with the flight-time 500~s. Let the initial number of $\nu _{1}$ neutrinos decrease in this flight by $500P_1=1/100$, hence $P_1=1/50000~s^{-1}$. This means that the half-life time is given by  $T_1=9.6$ h and $f=10.8$, as shown in Table~\ref{Table2}. The decreasing rate 1/100 of solar neutrinos may be likely in the error of the observable values \cite{ref:Mikheyev}. \\
\indent As a result, there is a possibility that no neutrinos with  $m_1$ exist in cosmic background neutrinos, but surviving neutrinos are those with heavier masses $m_2$ and $m_3$.   
All cases of $f=$10.8-41 may be also likely for solar neutrinos and for SN1987A-neutrinos. For the special case of  $f=$41 all of $\nu _1$, $\nu_2$, and $\nu_3$ appear to be stable in our universe, though we are not interested in such a case. 
%
\section{Dispersion relations} \label{sec:4}
%
 Let us consider two points A and B with a distance $L_{AB}$ on the $D3$-brane.  A neutrino can move directly 
on the $D3$-brane from A to B without tunneling. The propagation time $T$ of such a neutrino is, of course, 
given by $T=L_{AB}/|\dot{\vec{r}}|$, where the neutrino velocity is given by, in a conventional notation
\begin{align}
 & |\dot{\vec{r}}| =\frac{|\vec{p}|}{E} =\frac{\sqrt{E^2-m^2}}{E} ~.
 \label{eq:neutrino_v}
\end{align}
\indent On the other hand, the neutrino can move through the bulk from A to B according to reflections at 
the potential walls. This means as follows: One neutrino has a tunneling probability $P$ through the potential 
from 0 to $y_1$, then it escapes into the bulk and will simply continue its propagation in the bulk, never 
coming back on the brane. The initial numbers of neutrinos will decrease because of escapes by tunneling. \\
\indent However, there are neutrinos which are reflected from both points of the potential walls, 0 and  $y_1$. 
The total reflection probability is, of course, given by $R=1-P$. Thus a neutrino has a probability such that 
it travels from A on the brane into the $y$-direction by tunneling and will come back to B on the brane 
after reflection at both points of the potential walls. \\
\indent The neutrino energy inside the potential is given by Eq.(2.11), or in a conventional notation,
\begin{align}
 & E =\sqrt{\vec{p}^{\ 2} + e^{-2\eta |y|}\big(p_y^2+m^2\big)}
 \nonumber \\
 &  =\sqrt{\vec{p}^{\ 2} + m^2 q}~.
 \label{eq:neutrino_E} 
\end{align}
where $p_y^{\ 2}<0$.  This can be regarded as a dispersion relation in the bulk.  From Eq.(\ref{eq:neutrino_E}), we have the group velocity of neutrinos on the $D3$-brane 
\begin{align}
 & |\dot{\vec{r}}_{\mbox{bulk}}| =\frac{\partial E}{\partial |\vec {p}|} 
=\frac{\sqrt{E^2 - m^2 q}}{E}~.
 \label{eq:v_bulk}
\end{align}
\indent Eqs.(\ref{eq:neutrino_v}) and (\ref{eq:v_bulk}) tell us that there are two groups of neutrinos, one with the velocity (\ref{eq:neutrino_v}) and the other with (\ref{eq:v_bulk}). However, the difference between both velocities is too small to distinguish, because of $m^2$, $q<<E$. \\ 
\indent In our model we see trivially that there are no neutrinos with velocities faster than light because of the Poincare invariant metric of Eq.(\ref{eq:metric}) or equivalently by Eq. (\ref{eq:v_bulk}), even if they take a shortcut from A to B through the bulk. 
%
\section{Concluding remarks} \label{sec:5}
%
We have derived the potential in the extra-dimension from the five-dimensional warped spacetime $AdS_5$. 
Initially the potential works well to confine ordinary leptons on the $D3$-brane. However, we have the tunneling effect through the potential. 
The tunneling probability (\ref{eq:TunnelingA3}) for heavy leptons depends sharply on $m$. We have listed 
tunneling probabilities for heavy leptons in Table~\ref{Table1}. Their values seem to be very small, actually regarded so 
as to be almost zeros. They are hard to move into the extra dimension by the tunneling effect because of their large masses, but move only the  $D3$-brane.  \\
\indent On the other hand neutrino masses may be too small, so that we have generally a non-zero tunneling probability. We have taken neutrino masses of three mass-eigenstates as $m_1=0.001~$eV/$c^{2}$, $m_2=0.01~$eV/$c^{2}$, and $m_3=0.05~$eV/$c^{2}$ \cite{ref:Mikheyev}. Initial numbers of neutrinos will be found to decrease because of escapes into the fifth dimension by tunneling. From the data of solar neutrinos \cite{ref:Mikheyev} we estimate the half-life times of neutrinos with masses $m_1$, $m_2$, and $m_3$, respectively. The lightest neutrino with $m_1$ is only tunneling away into the extra dimension. Hence there is a possibility that no neutrinos with $m_1$ exist in cosmic background neutrinos, but surviving neutrinos are those with heavier masses  $m_2$ and $m_3$.\\
\indent All cases of $f=$10.8-41 in Table~\ref{Table2} may be also likely for solar neutrinos and for SN1987A-neutrinos. For the special case of $f=41$ all of $\nu_1$, $\nu_2$, and $\nu_3$ appear to be stable in our universe, though we are not interested in such a case. 
%
\begin{acknowledgments}
\indent We thank T. Okamura for many helpful discussions with us. Thanks are also due to referees for their appropriate advices, from which we have obtained the important result of dispersion relations in the bulk.
\end{acknowledgments}
%
%


\begin{thebibliography}{99}
\bibitem{ref:Photon_ex}
As for the photon, we can take into account of a fact that it runs on geodesic line 
$\dot{\vec{r}}^{\ 2}+\exp{(2k|y|)}\dot{y}^2=1$, which is derived from Eq.(\ref{eq:metric}). From this we have $\dot{y}=0$  whenever $|\dot{\vec{r}}|=1$, that is, the photon on the $D3$-brane never deviate into the extra dimension.
\bibitem{ref:mass_no_dependence}
The equation of motion for $y$ is derived from Eq.(2.12) as 
$\ddot{y}=m^2\eta [2\exp{(-4\eta y)}-q\exp{(-2\eta y)}]/c_o^{\ 2}$, for $y>0$, where 
$c_0^{\ 2}/m^2 = \vec {c}^2/m^2+\exp{(-2\eta y)}(1+p_y^{\ 2}/m^2)$ .From Eqs.(2.5) and (2.6) it follows that 
both $\vec{c}_0^{\ 2}/m^2$ and $p_y^{\ 2}/m^2$ are independent of $m$. Hence we see that $m^2/c_0^{\ 2}$ 
and $q$ are independent of $m$. This means that the acceleration $\ddot{y}$ is independent of any mass, 
reflecting the equivalence principle of General Relativity.
\end{thebibliography}
\end{document}